\documentclass[a4paper,DIV=14]{scrartcl}

\usepackage[bookmarks,
            raiselinks,
            pageanchor,
            hyperindex,
            colorlinks,
            citecolor=black,
            linkcolor=black,
            urlcolor=black,
            filecolor=black,
            menucolor=black]{hyperref}
\usepackage{graphicx}
\usepackage{microtype} 
\usepackage{authblk} 
\usepackage{amsmath, amssymb}
\usepackage[numbers]{natbib}    
\usepackage{listingscustomcpp} 

\hypersetup{pdftitle={Extending DUNE: The dune-xt modules},%
            pdfauthor={T. Leibner, R. Milk, F. Schindler}}

\lstnewenvironment{c++}[1][]{\lstset{style=cpp,#1}}{}
\newcommand{\mc}[1]{\lstinline!#1!}

\newcommand{\DUNE}{DUNE}
\newcommand{\dune}[1]{\texttt{dune-#1}}
\newcommand{\DXT}{\texttt{dune-xt}}
\newcommand{\DXTC}{\texttt{dune-xt-common}}
\newcommand{\DXTG}{\texttt{dune-xt-grid}}
\newcommand{\DXTL}{\texttt{dune-xt-la}}
\newcommand{\DXTF}{\texttt{dune-\allowbreak xt-\allowbreak functions}}
\newcommand{\abs}[1]{\left| #1 \right|}
\newcommand{\Python}{\textit{Python}}
\newcommand{\R}{\mathbb{R}}
\newcommand{\mcmath}[1]{\texttt{#1}}
\newcommand{\mcfoot}[1]{\footnote{\mc{#1}}}
\newcommand{\element}{t}
\newcommand{\grid}{\tau_h}

\newcommand{\eigen}{\mc{EIGEN}}
\newcommand{\N}{\mathbb{N}}
\newcommand{\Cpp}{C\raise .6ex \hbox{$_{++}$}}

\title{Extending \DUNE{}: The \DXT{} modules}
\author{Tobias Leibner}
\author{Ren\'e Milk\thanks{The author gratefully acknowledges funding by the DFG SPP 1648 ``Software
for Exascale Computing'' under contract OH 98/5-1.}}
\author{Felix Schindler}
\affil{Institute for Computational and Applied Mathematics, University of M\"unster,
Orleans-Ring 10, D-48149 M\"unster}

\begin{document}

\maketitle

\begin{abstract}
  We present our effort to extend and complement the core modules of the \emph{Distributed and Unified Numerics Environment} \DUNE{} (\url{http://dune-project.org}) by a well tested and structured collection of utilities and concepts.
  We describe key elements of our four modules \DXTC{}, \DXTG{}, \DXTL{} and \DXTF{}, which aim at further enabling the programming of generic algorithms within \DUNE{} as well as adding an extra layer of usability and convenience.
\end{abstract}

\section{Introduction}

Over the past decades, numerical approximations of solutions of partial differential equations (PDEs) have become increasingly important throughout almost all areas of the natural sciences.
At the same time, research efforts in the field of numerical analysis, computer science and approximation theory have led to adaptive algorithms to produce such approximations in an efficient and accurate manner: by now, for a large variety of PDEs advanced approximation techniques are available, such as wavelet methods \citep{Urb2009}, spectral methods \citep{GO1977}, radial basis functions \citep{Buh2000} and in particular grid-based discretization techniques such as Finite Volume and continuous or discontinuous Galerkin methods \citep{Cia1978}.

Focusing on the latter class, there exist a variety of open source and freely available PDE software frameworks with a strong scientific background, such as deal.ii \citep{BHH+2015}, \DUNE{} \citep{BBDEKO08A,BBDEKO08B}, Feel++ \citep{PCD+2012}, FreeFem++ \citep{Hec2012}, Fenics \citep{LMW2012} and libMesh \citep{KPSC2006}.
We consider the \Cpp{}-based \emph{Distributed and Unified Numerics Environment} (\DUNE{}), which consists of the core modules \dune{common}, \dune{geometry}, \dune{grid}, \dune{istl} \citep{BlaBas07,BlaBas08} and \dune{localfunctions}\footnote{All available under \url{http://dune-project.org}.}, complemented by the discretization frameworks \dune{fem} \citep{dunefem1,dunefem2}, \dune{gdt}\footnote{\url{https://github.com/dune-community/dune-gdt}} and \dune{pdelab} \citep{bastian2010generic}.

\DUNE{} yields highly efficient programs, is fairly well documented and has a large developer and user base with a strong background in numerical analysis and scientific computing.
However, since it is mainly targeted at researchers and centered around code efficiency, it has a steep learning curve and lacks some convenience features.
In addition, despite enabling flexible numerical schemes and providing an exceptionally powerful and generic grid concept, it does not allow for generic algorithms at all times, as detailed further below.
To remedy this situation we propose a family of \DUNE{} e\underline{xt}ension modules.

This paper is organized as follows: Section \ref{sec:intro} gives a brief overview of \DXT{} and presents our testing framework which is used in all modules.
The remaining sections each give an in-depth discussion of one of the \DUNE{} modules which make up \DXT{}.

\section{The \DXT{} project}
\label{sec:intro}
The \DXT{} project aims at providing an extra layer of usability and convenience to existing \DUNE{} modules and at the same time to enable programming of generic algorithms, in particular in the context of discretization frameworks.
It consists of the four modules \dune{xt-common}, \dune{xt-grid}, \dune{xt-la} and \dune{xt-functions} (which we refer to as \DXT{}), which are detailed throughout this paper.

We complement our discussion with extensive code listings and examples, which are not necessarily meant to be directly usable (for instance, we shall omit the \mc{main()} in \Cpp{} code and frequently drop the \mc{Dune::} namespace qualifier to improve readability).
In addition we provide references to code locations, where ``\mc{dune/foo/bar.hh}'' denotes the location of a file within the \dune{foo} module and ``\mc{dune/xt/foo/bar.hh}'' denotes the location of a file within the \dune{xt-foo} module.
However, we do not provide individual references for elements of the \Cpp{} standard template library, which are prefixed by \mc{std::}.\footnote{We refer to \url{http://en.cppreference.com/} instead.}

Since the main goal of \DXT{} is to provide a solid infrastructure, it does not ship examples for the usage of all of its parts.
Thus, we often refer to the generic discretization toolbox \dune{gdt}\footnote{\url{https://github.com/dune-community/dune-gdt}} for examples and motivation.

\subsection{Code availability}

The \dune{xt} modules are open source software and freely available on GitHub: \url{https://github.com/dune-community/}.\footnote{See Section \ref{sec:testing} on why we cannot make use of \url{https://gitlab.dune-project.org/}.}
They are mainly developed by T.~Leibner, R.~Milk and F.~Schindler with contributions from A.~Buhr, S.~Girke, S.~Kaulmann, B.~Verf\"urth and K.~Weber, amounting to roughly $43.000$ lines of code at the time of writing.
All modules are dual licensed under a BSD 2-Clause License\footnote{\url{http://opensource.org/licenses/BSD-2-Clause}} or any GPL-2.0+ License\footnote{\url{http://opensource.org/licenses/gpl-license}} with linking exception\footnote{\url{http://www.dune-project.org/license.html}}, thus being license-compatible to \DUNE{}.
We refer to the supplementary material, which contains a snapshot of the \dune{xt-super}\footnote{\url{https://github.com/dune-community/dune-xt-super}} module which bundles together all required dependencies and the \DXT{} modules described in this paper.

\subsection{Testing}
\label{sec:testing}
Testing code and exposing it to as many different circumstances as possible is an extremely important part of software development, especially given the templated nature of \DUNE{} and the large amount of provided grid managers.
In order to actually take a burden off of the developer, tests have to be executed automatically in a reliable manner on each published code change, should be carried out in as many different configurations as possible and reports of test failure have to be delivered to the developer responsible for the change.
In this section we describe the test procedure that is used in \DXT{}.

For testing, we use the Google \Cpp{} Testing Framework\footnote{\url{https://github.com/google/googletest}} (Gtest) in combination with the \DUNE{} module
\dune{testtools}\footnote{\url{https://gitlab.dune-project.org/quality/dune-testtools}}, introduced in \citep{Kempf2016}.
Further, we use Travis CI\footnote{\url{https://travis-ci.org/}} to automatically run tests on each push to the code repository.

Gtest is a testing library for \Cpp{} code of the xUnit family \citep{meszaros2007xunit}. It provides assertion
macros that make it very easy to write tests. Assertions can be non-fatal, such that the test will continue to run and output failure information if an assertion fails. This allows for the detection of several faults in one test cycle. Tests run independently of each other and, thanks to fixtures, objects for each test are created only for that specific test and are not shared between tests, which prevents hard-to-reproduce failures due to interacting tests. Fixture classes allow to use the same configuration for several different tests and to share code between tests, minimizing the effort needed for writing and maintaining tests.

We often want to run the same test for several related (template) classes. For this purpose, Gtest supports type parameterized test fixtures. The desired types have to be collected in the \mc{testing::Types} struct and passed to a test macro that automatically runs the test for all types. However, constructing a large \mc{Types} struct with permutations or products of type tuples is inconvenient. Moreover, Gtest currently only supports up to 50 types per test\footnote{\href{https://github.com/google/googletest/blob/master/googletest/include/gtest/internal/gtest-type-util.h}{\UrlFont https://github.com/google/googletest/blob/master/googletest/include/gtest/internal/\\gtest-type-util.h}} which is frequently exceeded in \DXT{}. Consider the following class from \DXTF{}, which represents a scalar-, vector-, or matrix-valued constant function $f: \Omega \to \R^{r \times c}$ that is localizable on each entity $t \in \tau_h$ of a grid view representing the domain $\Omega \subset \R^{\text{d}}$ (see Section \ref{sec:functions} for the notation of localizable functions and for an explanation of the template parameters of \mc{ConstantFunction}).
\begin{c++}
template <class E, class D, size_t d, class R, size_t r, size_t rC>
class ConstantFunction;
\end{c++}
Suppose that there existed a grid manager which was parametrized by the dimension of the grid and the type of the reference element (modeling a cube or a simplex geometry)\footnote{Only for the sake of presentation, the actual situation may of course be simpler or even more complex.}
\begin{c++}
template <int dim, class Geometry>
GridImplementation;
\end{c++}
and suppose we want to test the \mc{ConstantFunction} class for all combinations of dimensions \mc{d}, \mc{r}, \mc{rC} $\in \{1, 2, 3\}$ and both \mc{Geometry} types.
This adds up to $2 \cdot 3^3 = 54$ different specialized \mc{ConstantFunction} classes that would have to be manually written in the test code. Each time we want to add another grid type or higher dimensions, a lot of specializations have to be manually added.
In addition, macros are necessary to circumvent the $50$ types limit of Gtest complicating the test code even more.

To circumvent these problems, we use the \dune{testtools} module which provides tools for system testing in \DUNE. Among other features, \dune{testtools} makes it easy to create tests with dynamic and static variations. The configuration is done by providing a meta ini file that contains the possible variations (see Listing \ref{listing::metaini}).

\pagebreak

\begin{lstlisting}[caption=Meta ini file for test configuration, label=listing::metaini]
d  = 1, 2, 3 | expand
r  = 1, 2, 3 | expand
rC = 1, 2, 3 | expand
geometry = cube, simplex                   | expand
grid = GridImplementation<{d}, {geometry}> | expand
E = {grid}::Codim<0>::Entity
D = {grid}::ctype

[__static]
FUNCTIONTYPE = ConstantFunction<{E}, {D}, {d}, double, {r}, {rC}>
\end{lstlisting}

The \mc{expand} command tells \dune{testtools} to create all possible values of the corresponding key, the curved brackets make \dune{testtools} paste the content of the embraced variable. \mc{FUNCTIONTYPE} can be used like a preprocessor define in the associated \Cpp{} test code. Provided the meta ini file and the \Cpp{} test file, \dune{testtools} creates $54$ executables which each test a single specialization of \mc{ConstantFunction}. Adding another grid type or higher dimensions is as simple as adding it to the meta ini file without modifying the test code at all. In addition, dynamic variations, i.e.\ parameters that are provided at runtime and not at compile time, can easily be added in a similar way, see \citep{Kempf2016}.

Besides the reduced code complexity, this setup also makes it easier to debug tests as one can see at a glance which types are failing and debug only the associated tests. Creating an executable per tested type instead of testing all types in one executable using the \mc{Types} struct greatly reduces memory consumption during compilation. As a downside we see increased total compile times. Depending on the available build infrastructure this can be offset by the greater
available parallelism in building the test, resulting in more but smaller binaries.

\begin{figure}
\centering
  \begin{minipage}[b]{0.49\textwidth}
    \includegraphics[width=\textwidth]{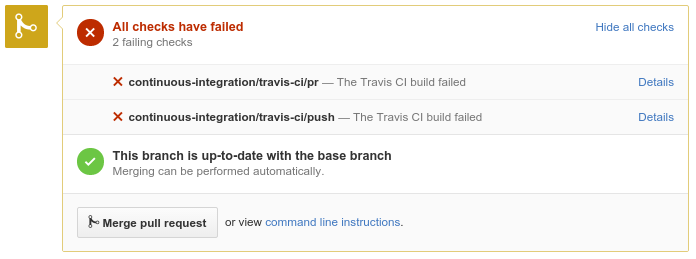}
  \end{minipage}
  \hfill
  \begin{minipage}[b]{0.49\textwidth}
    \includegraphics[width=\textwidth]{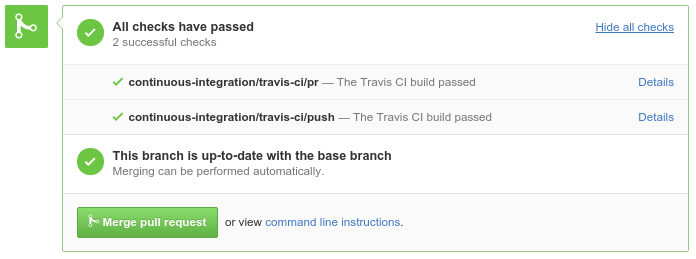}
  \end{minipage}
\caption{Travis integration with Github gives visual feedback for automatic tests (failing: left, succeeding: right), in particular to keep the workload for library developers at a minimum when handling code changes and contributions.}
\label{fig::travis_gh_pull}
\end{figure}

Writing the tests is only the first step. Even the most comprehensive test suite is useless without getting regularly executed.
We therefore continuously test each module on every update of the module's code repository using the Travis CI infrastructure
(Travis)\footnote{\url{https://travis-ci.org/dune-community}}, whose use is free of charge for open source projects. Relying on Google's Compute Engine\footnote{\url{https://cloud.google.com/compute/}} and following the setup specified in the module's \mc{.travis.yml} file Travis provides
us with a highly flexible environment to run our test suites in. Using Travis' ``build matrix``\footnote{\url{https://docs.travis-ci.com/user/customizing-the-build/\#Build-Matrix}} feature our test suites
are run with a diverse setup of available \DUNE{} modules, from the minimal set of hard dependencies specified in each
\mc{dune.module} file to the full set of suggested ones, as well as different compilers (currently gcc 4.9 and 5.3 and clang 3.7). Travis offers
tight integration with Github\footnote{\url{https://github.com}} pull-requests (see Figure \ref{fig::travis_gh_pull}) and branches, each of which gets automatically tested
whenever new commits get added, with visual on-page and email feedback. This is one of the main reasons why we host our modules on
Github. In our opinion, the prompt and automatic feedback on the validity of code contributions, with reasonably high
coverage, is of enormous value to the development process.

\section{The \DXTC{} module}
\label{sec:common}

Just like \dune{common}, we provide a common infrastructure in \dune{xt-common} with minimal dependencies, which is used throughout the other modules.

\subsection{Improved handling of dense containers}
\label{subsection::vectormatrixabstraction}

\dune{common} contains the
\mc{Field}\-\mc{Vector}\mcfoot{dune/common/fvector.hh} and
\mc{Field}\-\mc{Matrix}\mcfoot{dune/common/fmatrix.hh} classes, which model
small dense vectors and matrices of fixed size (in particular used for
coordinates, affine reference maps and function evaluations).
These containers are implemented to provide maximum performance but lack some
convenience features.
In particular, the lack of certain operators and constructors makes it difficult to
write generic code.
Consider, for instance, a generic string conversion utility (assuming \mc{T} is
a matrix type):
\begin{c++}
template <class T>
static inline T from_string(const std::string ss,
                            const size_t rows = 0,
                            const size_t cols = 0)
{
  T result(rows, cols); // <- does not compile for FieldMatrix
  // ... fill result from ss
  return result;
}
\end{c++}
The above example does compile if \mc{T} is a
\mc{Dynamic}\-\mc{Matrix}\mcfoot{dune/common/dynmatrix.hh} but not if \mc{T} is
a \mc{Field}\-\mc{Matrix}, which makes it extremely difficult to write generic
code.\footnote{%
  While it is clear that constructing a \mc{Field}\-\mc{Matrix} of fixed size
\mc{M}$ \times $\mc{N} is not sensible for other values of \mc{rows} and
\mc{cols}, the construction in line 6 should be possible for \mc{rows = M} and
\mc{cols = N} (throwing an appropriate exception otherwise).
}

In order to allow for generic algorithms we provide templated
\mc{Vector}\-\mc{Abstraction} and \mc{Matrix}\-\mc{Abstraction} classes in
\mc{dune/xt/common/\{matrix,vector\}.hh} (along with specialization for all
sensible vector and matrix classes), which allow for generic creation of and
access to matrices and vectors.
Additionally, we provide \mc{is_vector} and \mc{is_matrix} traits, which allow
to rewrite the above example:

\begin{c++}
#include <dune/xt/common/matrix.hh>
#include <dune/xt/common/type_traits.hh>

using namespace Dune::XT::Common;

template <class T>
  static inline typename std::enable_if<is_matrix<T>::value, T>::value
from_string(const std::string ss, const size_t rows = 0, const size_t cols = 0)
{
  auto result = MatrixAbstraction<T>::create(rows, cols);
  // ... fill result from ss using MatrixAbstraction< T >::set_entry(...)
  return result;
}
\end{c++}
It is thus possible to use \mc{from_string} with matrices of different
type:
\begin{c++}
#include <dune/common/fmatrix.hh>
#include <dune/common/dynmatrix.hh>

auto fmat = from_string<Dune::FieldMatrix<double, 2, 2>>("[1. 2.; 3. 4.]");
auto dmat = from_string<Dune::DynamicMatrix<double>>(    "[1. 2.; 3. 4.]");
\end{c++}
In particular, one can use it with any custom matrix implementation by providing
a specialization of \mc{Matrix}\-\mc{Abstraction} within the user code:
\begin{c++}
class CustomMatrix { /* user provided matrix implementation */ };

struct MatrixAbstraction<CustomMatrix> { /* implement specialization */ };

auto cmat = from_string<CustomMatrix>("[1. 2.; 3. 4.]");
\end{c++}

Based on these abstractions we provide many generic implementations in \dune{xt}.
For instance, we provide an extension of the
\mc{Float}\-\mc{Cmp}\mcfoot{dune/common/float_cmp.hh} mechanism from
\dune{common} (see Section \ref{sec:floatcmp}) for any combination of vectors (which allows for the same syntax
as its counterpart in \dune{common}, including compare styles and tolerances, see Section \ref{sec:floatcmp}):
\begin{c++}
#include <dune/common/dynvector.hh>
#include <dune/xt/common/float_cmp.hh>

std::vector< double >         svector({1., 1.});
Dune::DynamicVector< double > dvector(2, 1.);

Dune::XT::Common::FloatCmp::eq(svector, dvector);
\end{c++}

\subsection{Improved string handling and \texorpdfstring{\mc{Configuration}}{Configuration}}

As already hinted at above, we provide the string conversion
utilities
\begin{c++}
template< class T >
static inline T from_string(const std::string ss,
                            const size_t size = 0, const size_t cols = 0);

template< class T >
static inline std::string to_string(const T& ss);
\end{c++}
in \mc{dune/xt/common/string.hh}.
These can be used with any basic type as well as with all matrices and vectors
supported by the abstractions from Section \ref{subsection::vectormatrixabstraction}.
We use standard notation for vectors (\mc{"[1 2]"}) and matrices (\mc{"[1 2; 3
4]"}), see the previous paragraph for examples.
The \mc{from_string} function takes optional arguments which determine the size
of the resulting container (for containers of dynamic size), where $0$ means
automatic detection (the default).

Based on these string conversion utilities, we provide an extension of
\dune{common}'s \mc{Parameter}\-\mc{Tree}\mcfoot{dune/common/parametertree.hh}
in \mc{dune/xt/common/configuration.hh}: the \mc{Configuration} class.
The \mc{Con}\-\mc{fig}\-\mc{u}\-\mc{ra}\-\mc{tion} is derived from \mc{Parameter}\-\mc{Tree} and can thus be
used in all places where a \mc{Parameter}\-\mc{Tree} is expected.
While it also adds an additional layer of checks (in particular regarding
provided defaults), report and serialization facilities, one of its main
features is to allow the user to extract any type that is supported by the
string conversion facilities.
Given a sample configuration file in \mc{.ini} format (for example the default
configuration required to create a cubic grid using the factory methods discussed below),
\begin{minipage}{\linewidth}
\begin{lstlisting}[label={listing::dunegdt::our::cube_provider_cfg},%
caption=Contents of the \mc{cube_gridprovider_default_config()} in \dune{xt-grid}]
type            = xt.grid.gridprovider.cube
lower_left      = [0 0 0 0]
upper_right     = [1 1 1 1]
num_elements    = [8 8 8 8]
num_refinements = 0
overlap         = [1 1 1 1]
\end{lstlisting}
\end{minipage}
we can query the resulting \mc{Configuration} object \mc{config} for the types
supported by the \mc{Parameter}\-\mc{Tree},
\begin{c++}
auto num_refinements = config.get<int>("num_refinements");
\end{c++}
as well as for all types supported by our string conversion utilities (including
custom matrix and vector types as explained in the previous paragraph):
\begin{c++}
auto lower_left = config.get<FieldVector<double, dimDomain>>("lower_left");
\end{c++}
The above is valid for all $0 \leq \,$\mc{dimDomain}$\, \leq 4$, due to the
automatic size detection of \mc{from_string}.

\subsection{Timings}

Getting information about how long certain portions of an application take to run is crucial to guide
optimization efforts. It can also be useful to estimate the entire loop execution time from the first couple of loops
and of course for benchmarking parts of or an entire application for different parameters. Mature and generally simple to use tools exist, both commercial (e.g., Intel VTune\footnote{\url{https://software.intel.com/en-us/intel-vtune-amplifier-xe}}, Allinea Map\footnote{\url{http://www.allinea.com/products/map}}) and free (e.g., Vagrind/Callgrind\footnote{\url{http://valgrind.org/info/tools.html\#callgrind}}, Oprofile\footnote{\url{http://oprofile.sourceforge.net/}}), that allow very fine-grained performance analysis of a given code, down to function or even instruction level. It is however very hard or impossible to measure
custom, arbitrary sections of code with these tools and to present those measurements in an easily understandable format.
The \dune{common} module provides a straightforward implementation of a \mc{Timer}\footnote{\mc{dune/common/timer.hh}} class that relies on \mc{std::system_clock}
to measure expired walltime.
This is enough for simple needs, but we extend on this concept with the \mc{Timings} and related classes in \mc{dune/xt/common/timings.hh},
because measuring user and system time can be greatly insightful and having a centralized managing facility with
its own output capabilities is much easier for the user. The global \mc{Timings}
instance keeps a map of named sections and associated \mc{TimingData} objects. The user can start/stop sections
manually or use a guard object to start a timing at object construction and stop it when the guard object goes out of scope:

\begin{lstlisting}[language=C++,style=cpp,caption={Timing example}]
using namespace Dune::XT::Common;
timings().start("sec");
for (auto i : value_range(5)) {
  ScopedTiming scoped_timing("sec.inner");
  busywait(i * 200);
}
timings().stop("sec");
busywait(1000);
auto file = make_ofstream("example.csv");
timings().output_all_measures(*file);
\end{lstlisting}

Internally, \mc{TimingData} expands on \mc{Timer} by measuring wall, system and user time between start and stop
concurrently using a \mc{boost::timer::cpu_timer}\footnote{\url{http://www.boost.org/doc/libs/1_58_0/libs/timer/doc/cpu_timers.html\#Class-cpu_timer}}. \mc{Timings} has a couple of \mc{output_*} functions that
all write data in the Comma Separated Values (CSV) format and which distinguish themselves from one another in which
measures are produced and whether averages, minima and maxima are calculated over MPI-ranks.
We chose the CSV format for our output because it enables post processing and analysis by a wide variety of tools, e.g. Pandas \citep{mckinney2010} or matplotlib \citep{Hunter2007}.

\begin{lstlisting}[caption={example.csv}]
threads,ranks,sec_avg_usr,sec_max_usr,sec_avg_wall,sec_max_wall, sec_avg_sys,sec_max_sys,sec.inner_avg_usr,sec.inner_max_usr, sec.inner_avg_wall,sec.inner_max_wall,sec.inner_avg_sys, sec.inner_max_sys
8,1,2050,2050,2050,2050,0,0,2050,2050,2050,2050,0,0
\end{lstlisting}

\subsection{Extending Float Comparison}
\label{sec:floatcmp}

Floating point operations are not always exact, so using \mc{operator==} to compare primitive floating point types may yield unwanted results. To avoid this problem, \dune{common} provides functions \mc{eq}, \mc{ne}, \mc{gt}, \mc{lt}, \mc{ge} and \mc{le} in the namespace \mc{Dune::FloatCmp} to approximately compare floating point numbers. There are three variants of approximate comparisons (\mc{absolute}, \mc{relativeWeak} and \mc{relativeStrong}) implemented in \dune{common}. The \mc{absolute} comparison checks two floating point numbers $a$ and $b$ for equality by
\[ 
\abs{a - b} \leq \epsilon_{\text{abs}},
\]
where $\epsilon_{\text{abs}}$ is a specified tolerance parameter. This works well for numbers that are neither too small nor too large. However, two numbers with absolute values less than $\epsilon_{\text{abs}}$ will always be specified as equal while for large numbers the gap between two adjacent floats may become greater than $\epsilon_{\text{abs}}$ rendering the result the same as with \mc{==}. The comparisons \mc{relativeWeak}
\[
\abs{a-b} \leq \epsilon_{\text{rel}} \max(\abs{a}, \abs{b})
\]
and \mc{relativeStrong}
\[
\abs{a-b} \leq \epsilon_{\text{rel}} \min(\abs{a}, \abs{b})
\]
use a tolerance parameter that is scaled with the numbers and thus are suitable for a wide range of numbers. However, problems occur if one of the numbers is zero. Thus, we provide a fourth compare style \mc{numpy} in \mc{dune/xt/common/float_cmp.hh} that is equivalent to the implementation in \mc{numpy.isclose}\footnote{\url{http://docs.scipy.org/doc/numpy-dev/reference/generated/numpy.isclose.html}}. Here, both an absolute tolerance $\epsilon_{\text{abs}}$ and a relative tolerance $\epsilon_{\text{rel}}$ are chosen and the numbers $a$, $b$ are considered equal if
\[
\abs{a-b} \leq \epsilon_{\text{abs}} + \epsilon_{\text{rel}} \abs{b}.
\]
This allows for an absolute comparison (with a sufficiently small $\epsilon_{\text{abs}}$) near $0$ and a relative comparison elsewhere and is thus the default compare style in \DXTC{}. Note that this comparison is not symmetric with respect to $a$, $b$, so \mc{XT::Common::FloatCmp::eq(a,b)} may not be the same as \mc{XT::Common::FloatCmp::eq(b,a)} in some cases. The \mc{absolute}, \mc{relativeWeak} and \mc{relativeStrong}  compare styles are also supported by \DXTC{} and can be chosen by providing a template parameter to the comparison functions.

Often, we rather want to compare vectors of floating point numbers instead of scalar values. This could be done by looping over the components of the vectors and comparing them one by one, which is inconvenient for the user. \DXTC{} has built-in support for vector comparisons, i.e.\ one can compare any two vectors as long as there are  \mc{VectorAbstraction}s (see Section \ref{subsection::vectormatrixabstraction}) available for both vector types. In particular, \mc{std::complex} and vectors containing \mc{std::complex} values are supported.

\section{The \DXTG{} module}

The \dune{xt-grid} module builds on \dune{grid} and \dune{xt-common} to provide powerful concepts to improve performance (such as the \mc{EntityInlevelSearch} and the \mc{Walker}) or to allow for generic algorithms (such as the \mc{GridProvider}).

\subsection{Searching the grid}
\label{sec:search}

A typical task within a PDE solver is to prolong discrete function data from one discrete function space onto another one (that is usually associated with a finer grid).
Without a direct, one-to-one identification mapping of degrees of freedom (DoF) available (think of different discrete functions stemming from \dune{fem} and \dune{pdelab}), this requires the identification of those entities in the source space where the source function should be evaluated in quadrature points defined with respect to the range space.
With the \mc{EntityInlevelSearch} we provide a means to that end in \mc{dune/xt/grid/search.hh} for arbitrary grids, as long as the range grid covers a subset of the physical domain of the source grid.
\begin{c++}
template <class GridViewType, int codim = 0>
class EntityInlevelSearch : public EntitySearchBase<GridViewType>
{
  // not all types shown
public:
  EntityInlevelSearch(const GridViewType& gridview)
    : gridview_(gridview)
    , it_last_(gridview_.template begin<codim>())
  {}

  template <class PointContainerType>
  EntityVectorType operator()(const PointContainerType& points);

private:
  const GridViewType gridview_;
  IteratorType it_last_;
};
\end{c++}
It is modeled after the
\mc{HierarchicSearch}\footnote{\mc{dune/grid/utility/hierarchicsearch.hh}} in \dune{grid}, with the extension
that instead of finding one \mc{Entity} for one point at a time we allow searching for a sequence of points (via \mc{operator()}) and return
a sequence of entities accordingly. As another improvement the implementation remembers the last search position on the source grid
across search calls. This greatly improves the performance over the naive implementation of restarting the search
for every \mc{operator()}, especially if the grids match.

\subsection{Providing generic access to grid views and grid parts}

There exists an unfortunate disagreement between \dune{grid} and \dune{fem},
whether grid views or grid parts are to be used to model a
collection of grid elements, which makes it hard to implement generic
algorithms.
Think of some code which needs to create a \mc{LevelGridView} or
\mc{LevelGridPart}, depending on the further use for a discrete function space
implemented via \dune{pdelab} or \dune{fem}:
\begin{c++}
#include <dune/fem/gridpart/levelgridpart.hh>

template <class GridType>
void create_level(GridType& grid, const int lv) {
  // either ?
  auto level_view = grid.levelGridView(lv);
  // or ?
  Dune::Fem::LevelGridPart<GridType> level_part(grid, lv);
  // ...
}
\end{c++}

In \mc{dune/xt/grid/gridprovider.hh} we thus provide the
\mc{GridProvider}, the purpose of which is to hold a grid object and to allow for a generic creation of grid views and grid parts to ultimately allow for generic code despite the above mentioned disagreement:
\begin{c++}
template <class GridType>
class GridProvider
{
  // not all methods and types shown ...
public:
  std::shared_ptr<GridType>& grid_ptr();
  GridType& grid();

  template <Layers layer_type, Backends backend>
  typename Layer<layer_type, backend>::Type layer(const int lv = 0);
};
\end{c++}
Together with the \mc{Layers} and \mc{Backends} enum
classes\mcfoot{dune/xt/grid/layers.hh}, this allows to write generic code by
passing a grid provider along with the required tag:
\begin{c++}
#include <dune/xt/grid/provider.hh>

using namespace Dune::XT::Grid;

template <class GridType, Backends backend>
void create_level(GridProvider<GridType>& grid_provider, const int lv) {
  auto level_part_or_view = grid_provider.layer<Layers::level, backend>(lv);
  // ...
}
\end{c++}

We provide several factory methods to create a \mc{GridProvider} (for
instance \mc{make_cube_grid}, which creates grids of rectangular domains and \mc{make_dgf_grid}
and \mc{make_gmsh_grid}, which either create grids from the respective definition file).
In addition, we provide a means to select a factory at runtime using the
\mc{GridProviderFactory} struct, given the type of the grid \mc{G} and a \mc{Configuration} object \mc{config}
(for instance using the default one for \mc{make_cube_grid}, see Listing
\ref{listing::dunegdt::our::cube_provider_cfg}),
\begin{c++}
#include <dune/xt/gridprovider.hh>

#using namespace Dune::XT::Grid;

auto grid_provider = GridProviderFactory<G>::create("xt.grid.gridprovider.cube",
                                                    config);
\end{c++}
which is particularly useful in conjunction with configuration files or \Python{}
bindings.

\subsection{Identification of domain boundaries}

PDE solvers need to prescribe
the solution's behaviour on the boundary of the computational domain.
In general, there may exist a large number of different boundary conditions in any given problem, and therefore any PDE solver needs to provide a way to identify parts of that domain boundary
and categorize them.
Unfortunately, \dune{grid} does not provide such a mechanism.

This problem is tackled in \dune{pdelab}, for instance, by deriving from
\mc{Dune::}\allowbreak\mc{TypeTree::}\allowbreak\mc{LeafNode}\footnote{\mc{dune/typetree/leafnode.hh}} and implementing the following methods:
\begin{c++}
virtual bool isDirichlet(const IntersectionGeometryType& intersection_geometry,
                         const FieldVector<D, d - 1>& coord) const;

virtual bool isNeumann(const IntersectionGeometryType& intersection_geometry,
                       const FieldVector<D, d - 1>& coord) const;
\end{c++}

This approach is however limited to the boundary types defined in \dune{pdelab} and the user has no means to extend this approach to other boundary types (one cannot just add a \mc{isCustomBoundary()} method to the interface).

We therefore propose a more flexible mechanism that is based on mapping the
boundary category to types derived from \mc{BoundaryType}\footnote{\mc{dune/xt/grid/boundaryinfo/types.hh}}.
\begin{c++}
class BoundaryType
{
protected:
  virtual std::string id() const = 0;

public:
  virtual bool operator==(const BoundaryType& other) const
  {
    return id() == other.id();
  }
};

class DirichletBoundary : public BoundaryType
{
protected:
  virtual std::string id() const override final
  {
    return "dirichlet boundary";
  }
};

class NeumannBoundary : public BoundaryType { /*...*/};

class RobinBoundary : public BoundaryType { /*...*/};
\end{c++}

For a given grid view (and its type of intersection), access to the type of the boundary is granted by the \mc{BoundaryInfo} class concept
\begin{c++}
template <class IntersectionType>
class BoundaryInfo
{
public:
  virtual BoundaryType& type(const IntersectionType& intersection) const = 0;
};
\end{c++} 
based on which a PDE solver library (such as \dune{gdt}) can provide generic algorithms which act only on parts of the
domain boundary by checking
\begin{c++}
if (boundary_info.type(intersection) == DirichletBoundary()) {
  // ...
} else if (boundary_info.type(intersection) == NeumannBoundary()) {
  // ....
} else
  // ...
\end{c++}

In addition, this concept can be easily extended by the user by simply deriving from \mc{BoundaryType}.
As implementations of the \mc{BoundaryInfo} concept, we currently provide:
\begin{itemize}
  \item
    \mc{AllDirichlet} and \mc{AllNeumann}, the purpose of which is
self-explanatory.
  \item
    \mc{NormalBased}, which allows to identify boundary intersections by the
direction of their outward pointing normal.
\end{itemize}

In order to allow for problem definition classes to define domain boundaries
independently of the type of the grid, we also provide the
\mc{Boundary}\-\mc{Info}\-\mc{Factory}.
User classes can hold a complete description of one of the boundary infos from above in a
\mc{Configuration} instance.
Given the type of the grid (and thus the type of an intersection), one is then
able to create an instance of one of the implementations of the
\mc{Boundary}\-\mc{Info} concept of correct type, as required:
\begin{c++}
#include <dune/grid/yaspgrid.hh>
#include <dune/grid/sgrid.hh>
#include <dune/xt/common/configuration.hh>
#include <dune/xt/grid/boundaryinfo.hh>

using namespace Dune::XT::Common;
using namespace Dune::XT::Grid;

class Problem
{
public:
  Configuration boundary_info_cfg()
  {
    Configuration config;
    config["type"]      = "xt.grid.boundaryinfo.normalbased";
    config["default"]   = "dirichlet";
    config["neumann.0"] = "[ 1. 0.]";
    config["neumann.1"] = "[-1. 0.]";
    return config;
  }
};

typedef typename Dune::YaspGrid<2>::LeafIntersection YI;
typedef typename Dune::SGrid<2, 2>::LeafIntersection SI;

Problem problem;
auto boundary_info_y = BoundaryInfoFactory<YI>::create(problem.boundary_info_cfg());
auto boundary_info_s = BoundaryInfoFactory<SI>::create(problem.boundary_info_cfg());
\end{c++}
The type of \mc{boundary_info_y}, for instance, is \mc{std::unique_ptr<BoundaryInfo<YI>>}.
Given a rectangular domain in $\R^2$, it models a Neumann boundary left and
right and a Dirichlet boundary everywhere else.

\subsection{Periodic Gridviews}

Periodic boundary conditions are frequently used to model an infinite domain. To apply periodic boundary conditions in the context of Finite Volume methods, intersections on the periodic boundary have to be linked to the intersection on the opposite side of the (finite) grid. For Finite Element methods, the indices of connected degrees of freedom (DoFs) on the boundary have to be identified and suitable constraints applied. In any case, the implementation can be time-consuming and cumbersome if there is no support from the underlying grid manager.

Support for periodic boundary conditions in \DUNE{} is varying from grid to grid. \mc{YaspGrid} supports periodic boundaries using the parallel \DUNE{} grid interface which requires the user to run a parallel program. Several grid managers including \dune{alugrid} \citep{ans23252} and \mc{SPGrid} \citep{martinnoltephd} support periodicity by gluing together edges of the unit cube \citep{martinnoltephd}. While this allows to obtain the neighboring entity of an intersection on the periodic boundary, DoFs still have to be modified by hand.

Regarding the uneven support for periodic boundary conditions in \DUNE{}, we provide the \mc{PeriodicGridView} class in \mc{dune/xt/grid/periodic_gridview.hh} which is derived from a given arbitrary \mc{Dune::GridView}.
As long as the \mc{GridView} models an axis-parallel hyperrectangle with conforming faces, it is enough to replace a \mc{GridView} by the corresponding \mc{PeriodicGridView} in existing code to apply periodic boundary conditions (see Listing \ref{listing:periodic}).
This is internally achieved by replacing the \mc{IndexSet} and iterators as well as the corresponding methods (\mc{begin}, \mc{end}, \mc{ibegin}, \mc{iend}, \mc{size} and \mc{indexSet}) of the \mc{GridView} by periodic variants, utilizing the search capabilities from Section \ref{sec:search}.
The remaining methods are forwarded to the underlying \mc{GridView}.
Periodic directions can be specified by supplying a \mc{std::bitset} (see Listing \ref{listing:periodic}). By default, all directions are made periodic.

\begin{c++}[caption=Usage of \mc{PeriodicGridView}, label=listing:periodic]
using namespace Dune::XT::Grid;
GridViewType grid_view = grid.leafGridView();
std::bitset<3> periodic_directions(std::string("100")); // periodic in z-direction
PeriodicGridView<GridViewType> periodic_grid_view(grid_view, periodic_directions);
// use periodic_grid_view from here on to apply periodic boundary conditions ...
\end{c++}

The \mc{PeriodicIntersectionIterator} that can be obtained by the \mc{ibegin} and \mc{iend} methods returns a \mc{PeriodicIntersection} when dereferenced. The \mc{PeriodicIntersection} behaves exactly like the non-periodic \mc{Intersection} except that it returns \mc{neighbor() == true} and an \mc{outside()} entity on the periodic boundary. The \mc{outside()} entity is the entity adjacent to the periodically equivalent intersection, i.e., the intersection at the same position on the opposite side of the domain. The \mc{indexSet()} method returns a \mc{PeriodicIndexSet} which assigns the same index to entities that are periodically equivalent. Hence, the \mc{PeriodicIndexSet} is usually smaller than the corresponding \mc{IndexSet}. The \mc{begin} and \mc{end} methods of \mc{PeriodicGridView} return a \mc{PeriodicIterator} which visits only one of several periodically equivalent entities.

Note that there existed a similar but independently developed class \mc{PeriodicGridPart} in \texttt{dune-fem} that contained a periodic index set but no periodic intersections.
Unfortunately, the \mc{PeriodicGridPart} was removed in the \texttt{dune-fem} release 1.2.

\subsection{Walking the grid}

Since we are considering grid-based numerical methods, we frequently need to
iterate over the elements $\element \in \grid$ of a grid view.
In order to minimize the amount of required grid iterations, we want to be able
to carry out $N$ operations on each grid element, as opposed to walking the entire grid $N$ times.
We thus provide in \mc{dune/xt/grid/walker.hh} the templated \mc{Walker}
class, working with any \mc{Grid}\-\mc{View} (from \dune{grid}) or
\mc{Grid}\-\mc{Part} (from \dune{fem}):\footnote{%
  We provide traits in \mc{dune/grid/\{entity,intersection\}.hh} to extract
required information from a \mc{Grid}\-\mc{View} or \mc{Grid}\-\mc{Part} in a
generic way (see the \mc{Codim0} functor example).
}
\begin{c++}
template <class GridViewType>
class Walker
{
  // not all methods and types shown ...
public:
  void add(Functor::@Codim0@<GridViewType>& functor/*, ... */);
  void add(Functor::@Codim1@<GridViewType>& functor/*, ... */);
  void add(Functor::@Codim0And1@<GridViewType>& functor/*, ... */);

  void walk(const bool use_tbb = false);
};
\end{c++}

The user can add an arbitrary amount of functors to the \mc{Walker}, all of
which are then locally executed on each grid element.
Each functor is derived from one of the virtual interfaces \mc{Functor::Codim0},
\mc{Functor::Codim1} or \mc{Functor::Codim0And1}, for instance
\begin{c++}
template <class GridViewType>
class @Codim0@
{
public:
  typedef typename XT::Grid::Entity<GridViewType>::Type EntityType;

  virtual void prepare() {}
  virtual void apply_local(const EntityType& entity) = 0;
  virtual void finalize() {}
};
\end{c++}
where \mc{prepare} (and \mc{finalize}) are called before (and after) iterating
over the grid, while \mc{apply_local} is called on each element of the grid.
Each \mc{add} method of the \mc{Walker} accepts an additional argument which
allows to select the elements and faces the functor will be applied on.
For instance, in the context of a discontinuous Galerkin discretization in \dune{gdt} we want to apply
local coupling operators 
on all inner faces of the grid and local boundary operators 
on all Dirichlet faces of the grid.
Presuming we were given suitable implementations of these local operators as
functors and a \mc{BoundaryInfo} object in the sense of the previous paragraph,
the following would realize just that:
\begin{c++}
#include <dune/xt/grid/walker.hh>

using namespace Dune::XT::Grid;

Walker<GV> walker(grid_view);
walker.add(coupling_operator, new ApplyOn::InnerIntersectionsPrimally<GV>());
walker.add(boundary_operator, new ApplyOn::DirichletIntersections<GV>(boundary_info));
// add more, if required...
walker.walk()
\end{c++}

Note that the \mc{walk} method allows to switch between a serial and a shared
memory parallel iteration over the grid, at runtime (via the \mc{use_tbb}
switch).
In particular, the user only has to provide implementations of the functors and
does not need to deal with any parallelization issues (or different types of grid
walkers, depending on the parallelization paradigm).

\section{The \DXTL{} module}

As discussed (Section \ref{subsection::vectormatrixabstraction}), \dune{common}
provides small dense vectors and matrices which are, e.g., used for coordinates.
In addition, any PDE solver requires large vectors and (usually sparse) matrices to
represent assembled functionals and operators stemming from the underlying PDE.
Linear algebra containers are a performance critical aspect of any
discretization framework and one usually does not want to be restricted to a single implementation: there
exist external backends which are well suited for serial and shared memory
parallel computations (such as \eigen \citep{eigenweb}) while others are more suited for
distributed memory parallel computations (such as \dune{istl}).
Neither is fitting for every purpose and we thus require a means to exchange the
implementation of matrices and vectors depending on the circumstances.
This calls for abstract interfaces for containers and linear solvers, which we provide within the
\dune{xt-la} module.

\subsection{Generic linear algebra containers}

We chose a combination of static and dynamic inheritance for these interfaces,
allowing for virtual function calls that act on the whole container (such as
\mc{scal}) while using the ``Curiously recurring template patterns'' (\mc{CRTP},
see \citep{Cop1995}) paradigm for methods that are called frequently (such as
access to individual elements in loops), allowing the compiler to optimize
performance critical calls (for instance by inlining).
We provide a thread-safe helper class \mc{CRTPInterface} in
\mc{dune/xt/common/crtp.hh} along with thread-safe \mc{CHECK_...} macros for
debugging.\footnote{The tools provided in \mc{dune/common/bartonnackmanifcheck.hh}
may not work properly in a shared memory parallel program.}

All matrices and vectors are derived from \mc{ContainerInterface}\footnote{\mc{dune/xt/la/container/container-interface.hh}}:

\begin{c++}
template <class Traits, class ScalarType = typename Traits::ScalarType>
class ContainerInterface
  : public CRTPInterface<ContainerInterface<Traits, ScalarType>, Traits>
{
  // not all methods and types shown ...
public:
  typedef typename Traits::derived_type derived_type;
  
  // Sample CRTP implementation:
  inline void scal(const ScalarType& alpha)          
  {                                                  
    // as_imp() from CRTPInterface performs a static_cast into derived_type.
    // Thus, scal() of the derived class is called.
    CHECK_AND_CALL_CRTP(this->as_imp().scal(alpha));
  }
  inline void axpy(const ScalarType& alpha, const derived_type& xx);
  inline derived_type copy() const;

  // Default implementation:
  // could be overriden by any derived class.
  virtual derived_type& operator*=(const ScalarType& alpha) 
  {                                                         
    scal(alpha);
    return this->as_imp();
  }
};
\end{c++}
The \mc{ContainerInterface} enforces just enough functionality to assemble a
linear combination of matrices or vectors, which is frequently required in the
context of model reduction (compare \citep{DBLP:journals/corr/MilkRS15}).
Given an affine decomposition of a parametric matrix or vector $B$, we need
to assemble $\sum_{q = 0}^{Q - 1} \theta_q B_q$ for given component containers $B_q$ and scalar coefficients $\theta_q$.
The following generic code will work for any matrix or vector type \mc{C}
derived from \mc{ContainerInterface}:\footnote{Note that due to COW and move semantics for all matrices and vectors in \dune{xt-la}, the only deep copy is actually done in line 11 (neither in line 10 nor in line 14). Note also the use of the \mc{is_container} traits in line 6 that we provide in
\mc{dune/la/container/container-interface.hh}.
  Together with \mc{std::enable_if}, this checks that \mc{C} is derived from
\mc{Container}\-\mc{Interface} at compile time.}

\begin{c++}
#include <dune/xt/la/container/container-interface.hh>

using namespace Dune::XT::LA;

template <class C>
    typename std::enable_if<is_container<C>::value, C>::type
assemble_lincomb(const std::vector<C>& components,
                 const std::vector<double>& coefficients)
{
  auto result = components[0].copy();
  result *= coefficients[0];
  for (size_t qq = 1; qq < components.size(); ++qq)
    result.axpy(coefficients[qq], components[qq]);
  return result;
}
\end{c++}

All containers in \dune{xt-la} are implemented with ``copy-on-write'' (COW) \citep[190-194]{meyers1995more}.
Implementing data sharing among entities with data duplication only occurring if an entity tries to modify the referenced data is a well established and common technique in Computer Science.
This pattern (also sometimes denoted as ``lazy copy'') typically incurs minimal runtime overhead for the required reference counting for the great benefit of being
able to pass around copies of an entity without immediate expensive memory copies.
Since all container implementations in \dune{xt-la} are
proxy classes that forward operations to the respective backend instances, we insert
the COW logic into the \mc{backend} call, which is then used throughout the class to access the backend (in addition to allowing the user to directly access the underlying backend):
\begin{c++}
BackendType& backend()
{
  ensure_uniqueness();
  return *backend_;
}
\end{c++}
The \mc{backend_} is simply a \mc{std::shared_ptr<BackendType>} on which the \mc{ensure_uniqueness} method can query the current status by a call to \mc{unique} and perform the deep copy, if required.

Based on \mc{Container}\-\mc{Interface} we provide the
\mc{Vector}\-\mc{Interface}\mcfoot{dune/xt/la/container/vector-interface.hh}
for dense vectors and the
\mc{Matrix}\-\mc{Interface}\mcfoot{dune/xt/la/container/matrix-interface.hh}
for dense and sparse matrices.
Each derived vector class has to implement the methods \mc{size},
\mc{add_to_entry}, \mc{set_entry} and \mc{get_entry_ref}, which allow to access
and change individual entries of the vector.
The interface provides default implementations for all relevant mathematical
operators, support for range-based for loops and many useful methods, such as
\mc{dot}, \mc{mean}, \mc{standard_deviation} and \mc{l2_norm}, just to name a
few.

Each derived matrix class has to implement \mc{rows}, \mc{cols},
\mc{add_to_entry}, \mc{set_entry} and \mc{get_entry} to allow for access to
individual entries, \mc{mv} for matrix/vector multiplication and \mc{clear_row},
\mc{clear_col}, \mc{unit_row} and \mc{unit_col}, which are required in the
context of solving PDEs with Dirichlet Constraints or pure Neumann problems.
Every matrix implementation (even a dense one) is constructible from the same type of sparsity
pattern (which we provide in \mc{dune/xt/la/container/pattern.hh}) and the
access methods \mc{..._entry} are only required to work on entries that are
contained in the pattern.
The interface provides several mathematical operators, norms and a means to
obtain a pruned matrix (where all entries close to zero are removed from the
pattern).

We also provide several vector and matrix implementations:
\begin{itemize}
  \item
    The \mc{Common}\-\mc{Dense}\-\mc{Vector} and
\mc{Common}\-\mc{Dense}\-\mc{Matrix} in \mc{dune/xt/la/container/common.hh},\hspace{-0.35 pt}
based on the \mc{Dynamic}\-\mc{Vector} and \mc{Dynamic}\-\mc{Matrix} from
\dune{common}.
    These are always available.
  \item
    The \mc{Eigen}\-\mc{Dense}\-\mc{Vector},
\mc{Eigen}\-\mc{Mapped}\-\mc{Dense}\-\mc{Vector},
\mc{Eigen}\-\mc{Dense}\-\mc{Matrix} and
\mc{Eigen}\-\mc{Row}\-\mc{Major}\-\mc{Sparse}\-\mc{Matrix} in
\mc{dune/xt/la/container/eigen.hh}, based on the \eigen{}
package (if available).
    The \mc{Eigen}\-\mc{Mapped}\-\mc{Dense}\-\mc{Vector} allows to wrap an
existing \mc{double*} array and the
\mc{Eigen}\-\mc{Row}\-\mc{Major}\-\mc{Sparse}\-\mc{Matrix} allows to wrap
existing matrices in standard CSR format, which allows to wrap other container implementation
(for instance in the context of \Python{} bindings).
  \item
    The \mc{IstlDenseVector} and \mc{IstlRowMajorSparseMatrix} in
\mc{dune/xt/la/con}\-\mc{tainer/istl.hh}, based on \dune{istl} (if
available).
\end{itemize}

To allow for generic algorithms we also provide the \mc{Backends} enum class along with the \mc{default_back}\-\mc{end}, \mc{default_sparse_backend} and \mc{default_dense_backend} defines (which are set depending on the build configuration).
These can be used in other libraries and user code to obtain suitable containers independently of the current build configuration.

Consider, for instance, an $L^2$ projection of a function $f: \Omega \to \R$ onto a
discontinuous Galerkin discrete function space defined on a grid view $\tau_h$ modeling $\Omega$.
Given local basis functions $\varphi_i^t$, $0 \leq i < I$ of such a space on each entity $\element \in \grid$,
the local DoF vector $\underline{f_h^t} \in \R^I$ of the
projected discrete function is given as the solution of the linear system
\begin{align}
  \label{eq::projection}
  \underline{{L^2}^t} \cdot \underline{f_h^t} = \underline{l_h^t},
  &&\text{ with }&&\big(\underline{{L^2}^t}\big)_{i, j} := \int_t \varphi_i^t \varphi_j^t &&\text{ and }&&\big(\underline{l_h^t}\big)_i := \int_t f \varphi_i^t.
\end{align}
The problem of assembling and solving \eqref{eq::projection} is a typical situation within any PDE solver.
In \dune{gdt}, for instance, \eqref{eq::projection} is roughly assembled as follows, given a local basis and an appropriate quadrature:
\begin{c++}[label=listing:local_vec]
#include <dune/xt/la/container.hh>

using namespace Dune::XT::LA;

typedef typename Container<double, default_dense_backend>::MatrixType LocalMatrixType;
typedef typename Container<double, default_dense_backend>::VectorType LocalVectorType;

// ... on each grid element
LocalMatrixType local_matrix(local_basis.size(), local_basis.size(), 0.);
LocalVectorType local_vector(local_basis.size(), 0.);
LocalVectorType local_DoFs(local_basis.size(), 0.);

// ... at each quadrature point, given evauations of the local basis as basis_values
//     and of the source function as source_value
for (size_t ii = 0; ii < local_basis.size(); ++ii) {
  local_vector[ii] += integration_element * quadrature_weight
                    * (source_value * basis_values[ii]);
  for (size_t jj = 0; jj < local_basis.size(); ++jj) {
    local_matrix.add_to_entry(ii, jj,
                              integration_element * quadrature_weight
                              * (basis_values[ii] * basis_values[jj]));
  }
}
\end{c++}
Note that we neither have to manually specify the correctly matching matrix and
vector types nor to include the correct headers (which depend on the current
build configuration).
The \mc{Container} traits together with any of the \mc{default_...} defines
always yields appropriate available types (lines 1, 5, 6).

\subsection{Runtime selectable solvers}

In order to determine the local DoF vector in the above example, we need to
solve the algebraic problem: find \mc{local_DoFs}, such that
\begin{align}
  \mcmath{local\_matrix} \cdot \mcmath{local\_DoFs} = \mcmath{local\_vector}.
\notag
\end{align}
In addition to such small, dense problems we also require the inversion of large
(sparse) system and product matrices.
For interesting large and real-world problems, however, there are few linear
solvers available which can be used as a black box (if at all).
Most problems require a careful choice and detailed configuration of the correct
linear solver.
We thus require access to linear solvers which can be used in a generic way but
also exchanged and configured at runtime.

In \mc{dune/xt/la/solver.hh} we provide such solvers via the \mc{Solver}
class:
\begin{c++}
template <class MatrixType>
class Solver
{
  // simplified variant
public:
  Solver(const MatrixType& matrix);

  static std::vector<std::string> types();

  static Configuration options(const std::string type = "");

  template <class RhsType, class SolutionType>
  void apply(const RhsType& rhs, SolutionType& solution) const;

  template <class RhsType, class SolutionType>
  void apply(const RhsType& rhs, SolutionType& solution,
             const std::string& type) const;

  template <class RhsType, class SolutionType>
  void apply(const RhsType& rhs, SolutionType& solution,
             const Configuration& options) const;
};
\end{c++}
We provide specializations of the \mc{Solver} class for all matrix
implementations derived from \mc{Matrix}\-\mc{Interface} (see the previous paragraph).
To continue the example from the previous paragraph, this allows to determine
the local DoF vector of the $L^2$ projection:
\begin{c++}
#include <dune/xt/common/exceptions.hh>
#include <dune/xt/la/solver.hh>

try {
  XT::LA::Solver<LocalMatrixType>(local_matrix).apply(local_vector, local_DoFs);
} catch (XT::Exceptions::linear_solver_failed& ee) {
  DUNE_THROW(Exceptions::projection_error,
             "L2 projection failed because a local matrix could not be inverted!\n\n"
             << "This was the original error: " << ee.what());
}
\end{c++}
The above example shows a typical situation within the library code of
\dune{gdt}: we need to solve a small dense system for provided matrices and
vectors of unknown type.
We can do so by instantiating a \mc{Solver} and calling the black-box variant of
\mc{apply} (line 5).
This apply variant is default implemented by calling
\begin{c++}
apply(rhs, solution, types()[0]);
\end{c++}
where \mc{types()} always returns a (non-empty) list of available linear solvers
for the given matrix type, in descending priority (meaning the first is supposed
to ``work best'').
For instance, if \mc{local_matrix} was an \mc{EigenDenseMatrix}, a call to
\mc{types()} would reveal the following available linear solvers:
\begin{lstlisting}
{"lu.partialpiv", "qr.householder", "llt", "ldlt", "qr.colpivhouseholder",
 "qr.fullpivhouseholder", "lu.fullpiv"}
\end{lstlisting}
On the other hand, if the matrix was an \mc{EigenRowMajorSparseMatrix}, a call
to \mc{types()} would yield
\begin{lstlisting}
{"bicgstab.ilut", "lu.sparse", "llt.simplicial", "ldlt.simplicial",
 "bicgstab.diagonal", "bicgstab.identity", "qr.sparse",
 "cg.diagonal.lower", "cg.diagonal.upper", "cg.identity.lower",
 "cg.identity.upper"}
\end{lstlisting}
Given a (large sparse) \mc{system_matrix} (for instance stemming from a
discretized Laplace operator) and \mc{rhs} vector, we can solve the
corresponding linear system using a specific solver by calling
\begin{c++}
#include <dune/xt/la/solver.hh>

XT::LA::Solver<SystemMatrixType> linear_solver(system_matrix);
linear_solver.apply(rhs, solution, "ldlt.simplicial");
\end{c++}
The above call to \mc{apply} is default implemented by calling
\begin{c++}
apply(rhs, solution, options(type));
\end{c++}
where \mc{options(type)} always returns a \mc{Configuration} object with
appropriate options for the selected \mc{type}.
With \mc{type = "ldlt.simplicial"}, for instance, we are implicitly using the
following options (which are the default for \mc{"ldlt.simplicial"}):
\begin{lstlisting}
type                     = ldlt.simplicial
post_check_solves_system = 1e-5
check_for_inf_nan        = 1
pre_check_symmetry       = 1e-8
\end{lstlisting}

All implemented solvers default to checking whether the computed solution does
actually solve the linear system and provide additional sanity checks.
We make extensive use of exceptions if any check is violated, which allows to
recover from undesirable situations in library code (as shown in the example
above).\footnote{%
  We also provide our own implementation of the \mc{DUNE_THROW} macro in
\mc{dune/xt/common/exceptions.hh} which replaces the macro from
\dune{common} and can be used with any \mc{Dune::Exception}.
  Apart from a different formatting (and colorized output), it also provides
information of interest in distributed memory parallel computations.
}
The \mc{"ldlt.simplicial"} solver, for instance, does only work for symmetric
matrices and we thus check the matrix for symmetry beforehand (which can be disabled by
setting \mc{"pre_check_symmetry"} to \mc{0}).
Each type of linear solver provides its own options, which allow the user to
fine-tune the linear solver to his needs.
For instance, the iterative \mc{"bicgstab"} solver with \mc{"ilut"}
preconditioning accepts the following options (in addition to
\mc{"post_check_solves_system"} and \mc{"check_for_inf_nan"}, which are always
supported):
\begin{c++}
#include <dune/xt/la/solver.hh>

XT::LA::Solver<SystemMatrixType> linear_solver(system_matrix);
auto options = linear_solver.options("bicgstab.ilut");

options["max_iter"]  = 1000;
options["precision"] = "1e-14";
options["preconditioner.fill_factor"] = 10;
options["preconditioner.drop_tol"]    = "1e-4";

linear_solver.apply(rhs, solution, options);
\end{c++}

The actually implemented variant of \mc{Solver} in \dune{xt-la} also takes a
communicator as an optional argument, to allow for distributed parallel linear
solvers.
We provide several implementations of such parallel solvers based on
\dune{istl}.
A linear solver for \mc{IstlRowMajorSparseMatrix}, for instance, supports the
following \mc{types()}, most of which can be readily used in distributed
parallel environments:
\begin{c++}
{
#if !HAVE_MPI && HAVE_SUPERLU
 "superlu",
#endif
            "bicgstab.amg.ilu0", "bicgstab.amg.ssor", "bicgstab.ilut",
 "bicgstab.ssor", "bicgstab"
#if HAVE_UMFPACK
                            , "umfpack"
#endif
}
\end{c++}

\section{The \DXTF{} module: Local and localizable functions}
\label{sec:functions}

A correct interpretation of data functions and a correct handling of analytical and discrete functions is an important part of every PDE solver.
Given a domain $\Omega \subset \R^d$ and a grid view $\tau_h$ of this domain, one is usually not interested in considering a function $f: \Omega \to \R$, but rather its \emph{local function} $f^t := f|_t \circ \Phi^t$ with respect to a grid element $t \in \tau_h$, where $\Phi^t: \hat{t} \to t$ denotes the reference mapping from the respective reference element $\hat{t}$ to the actual element $t$, as provided by \dune{grid}.\footnote{We only present the scalar case here. The same concepts apply to vector- and matrix-valued functions as well, as implemented in \dune{xt-functions}.}
This allows to compute integrals in terms of quadratures from \dune{geometry} and shape functions from \dune{localfunctions} defined on the reference elements of a grid.
In addition, one is interested in ``localized derivatives'' of $f$, for instance its \emph{localized gradient} $\nabla_t f_t := \nabla f \circ \Phi^t$.
While the localized gradient of a function does not coincide with the gradient of a local function it is an important tool to preserve the structure of an integrand, when transformed to the reference elements for integration.\footnote{Using the chain rule, we obtain for the latter $\nabla f^t = \nabla( f \circ \Phi^t ) = ( \nabla f \circ \Phi^t) \nabla \Phi^t$, while the following holds for the former: $\nabla_t f^t = \nabla f \circ \Phi^t = \nabla(f \circ \Phi^t) (\nabla \Phi^t)^{-1} = \nabla f^t (\nabla \Phi^t)^{-1}$.}

Consider for instance the following integral arising in the weak formulation of a Laplace operator, which is transformed to the reference element $\hat{t}$, where $a$, $\varphi$ and $\psi$ denote scalar functions and $\Lambda_t^2 := |\det(\nabla {\Phi^t}^\bot \nabla \Phi^t)|$:
\begin{align}
  \int_t (a \nabla \psi) \cdot \nabla \varphi
    &= \int_{\hat{t}} \Lambda_t\, \big( (a \circ \Phi^t)(\nabla \psi \circ \Phi^t) \big) \cdot (\nabla \varphi \circ \Phi^t).
\notag
\intertext{Using the definition of a local function and a localized gradient, the above is equivalent to}
    \dots &= \int_{\hat{t}} \Lambda_t\, (a^t \nabla_t \psi^t) \cdot \nabla \varphi^t,
\label{eq:laplace}
\end{align}
which nicely preserves the structure of the original integrand.

For a given grid view $\grid$, we thus call a function \emph{localizable} with respect to $\grid$, if there exists a local function with localized derivatives for each element of the grid view.
These functions form the discontinuous space of locally polynomial functions with varying polynomial degree,
\begin{align*}
  Q(\grid) := \big\{\; q: \Omega \to \R \;\big|\; \forall \element \in \grid\; \exists k(\element) \in \mathbb{N}\text{, such that } q^t \in \hat{\mathbb{P}}_{k(t)}(t) \;\big\},
\end{align*}
where $\hat{\mathbb{P}}_k(t)$ denotes the set of polynomials $q: \hat{t} \to \R$ of maximal order $k$.
Note that in the context of integration, the polynomial degree of the integrand is limited by the availability of suitable quadratures and always finite in practical computations.
Thus, even functions such as $f(x) = sin(x)$ can in practical computations only be represented by surrogates $f \approx \tilde{f} \in Q(\grid)$.
Consequently, the space $Q(\grid)$ presents a common space for all functions, including analytically given data functions and, most importantly, discrete functions of any discretization framework.

This concept allows for a generic discretization framework such as \dune{gdt}, where all operators, functionals, projections and prolongations are implemented in terms of localizable functions.
Thus, a Laplace operator which locally realizes \eqref{eq:laplace} can be used to assemble a system or product matrix, where $\psi^t$ and $\varphi^t$ are basis functions of a discrete function space, or to compute the norm of any combination of analytical or discrete functions, in which case $\psi^t$ and $\varphi^t$ model the corresponding local functions.
In \dune{xt-functions}, we provide interfaces and implementations to realize this concept of \emph{localizable functions} and \emph{local functions} (not to be confused with shape functions from \dune{localfunctions}) and for the remainder of this section we discuss the realization of these concepts in \dune{xt-functions}.
A similar effort has been undertaken in the \dune{functions} module \citep{DBLP:journals/corr/EngwerGMS15}, independently of our work.\footnote{%
Since both efforts share the same mathematical basis it is planned to make \dune{xt-functions} compatible to \dune{functions} in future work.}

Given a grid element $\element \in \grid$, we model a collection of (scalar-, vector-
or matrix-valued) local functions $\varphi^\element: \hat{\element} \to \R^{r
\times c}$, for $r, c \in \N$ (where $\hat{\element}$ denotes the reference element associated with $\element$) by the \mc{LocalfunctionSetInterface} in
\mc{dune/xt/functions/interfaces.hh}:
\begin{c++}
template<class EntityType,
         class DomainFieldType, size_t dimDomain,
         class RangeFieldType,  size_t dimRange, size_t dimRangeCols = 1>
class LocalfunctionSetInterface
{
  // not all methods and types show ...
public:
  virtual const EntityType& entity() const

  virtual size_t size() const = 0;
  virtual size_t order() const = 0;

  virtual void evaluate(const DomainType& xx, std::vector<RangeType>& ret) const = 0;
  virtual void jacobian(const DomainType& xx,
                        std::vector<JacobianRangeType>& ret) const = 0;
};
\end{c++}
The template parameter \mc{Entity}\-\mc{Type} models the type of the grid
element $\element \in \grid$, the parameters \mc{Domain}\-\mc{Field}\-\mc{Type}
and \mc{dim}\-\mc{Domain} model $\R^d \supset t$ while
\mc{Range}\-\mc{Field}\-\mc{Type}, \mc{dim}\-\mc{Range} and
\mc{dim}\-\mc{Range}\-\mc{Cols} model $\R^{r \times c}$.\footnote{%
As a shorthand, we write \mc{<E, D, d, R, r, rC>} instead of \mc{<EntityType, DomainFieldType, dimDomain, RangeFieldType, dimRange, dimRangeCols>}.}
The resulting \mc{Domain}\-\mc{Type} is a \mc{FieldVector<DomainFieldType, dimDomaim>}, while \mc{Range}\-\mc{Type} and
\mc{Jacobian}\-\mc{Range}\-\mc{Type} are composed of \mc{Field}\-\mc{Vector} and
\mc{Field}\-\mc{Matrix}, depending on the dimensions.
Each set of local functions has to report its polynomial order and size.
The methods \mc{evaluate} and \mc{jacobian} expect vectors of size \mc{size()}
for \mc{ret}.

Note that we use a combination of static polymorphism, to fix the type of the grid and all dimensions at compile time, and dynamic polymorphism, which allows to exchange functions of same grid and dimensions at runtime.

By implementing all operators and functionals in terms of \mc{Localfunction}\-\mc{Set}\-\mc{Interface}, a discretization framework can realize the above claim of
unified handling of analytical and discrete functions.
This is for instance the case in \dune{gdt}, where the local bases of discrete function spaces are realized as implementations of \mc{Localfunction}\-\mc{Set}\-\mc{Interface}.
In addition, we also provide an interface for individual local functions, which can be used by data functions and
local functions of discrete functions:
\begin{c++}
template <class E, class D, size_t d, class R, size_t r, size_t rC>
class LocalfunctionInterface
  : public LocalfunctionSetInterface<E, D, d, R, r, rC>
{
  // not all methods and types shown ...
public:
  virtual void evaluate(const DomainType& xx, RangeType& ret) const = 0;
  virtual void jacobian(const DomainType& xx, JacobianRangeType& ret) const = 0;

  virtual size_t size() const override final
  {
    return 1;
  }
}
\end{c++}

Alongside, we also provide the following interface for \emph{localizable functions}
which mostly act as containers of local functions:
\begin{c++}
template <class E, class D, size_t d, class R, size_t r, size_t rC>
class LocalizableFunctionInterface
{
  // not all methods and types shown ...
public:
  virtual std::string name() const;

      virtual std::unique_ptr< LocalfunctionInterface<E, D, d, R, r, rC>>
  local_function(const EntityType& entity) const = 0;
};
\end{c++}
Based on this interface we provide visualization and convenience operators,
allowing for
\begin{c++}
(f - f_h).visualize(grid_view, "difference");
\end{c++}
where \mc{f} may denote a localizable data function and \mc{f_h} might denote
a discrete function, if both are localizable with respect to
the same \mc{grid_view}.
Expressions such as \mc{f - f_h}, \mc{f + f_h} or \mc{f*f_h} yield localizable
functions via generically implemented local functions (if the dimensions allow
it).

We also provide numerous implementations of \mc{LocalizableFunctionInterface} in
\DXTF, most of which can also be created using the
\mc{FunctionsFactory}\mcfoot{dune/xt/functions/factory.hh}, given a \mc{Configuration}:
\begin{itemize}
  \item
    \mc{CheckerboardFunction} in \mc{dune/xt/functions/checkerboard.hh}
models a piecewise constant function, the values of which are associated with an
equidistant regular partition of a domain.
    Sample configuration for a function $\R^2 \to \R$:
    \begin{lstlisting}
lower_left   = [0. 0.]
upper_right  = [1. 1.]
num_elements = [2 2]
values       = [1. 2. 3. 4.]
    \end{lstlisting}
  \item
    \mc{ConstantFunction} in \mc{dune/xt/functions/constant.hh} models a
constant function.
    Sample configuration for a function $\R^d \to \R^{2 \times 2}$, for any $d
\in \N$, mapping to the unit matrix in $\R^2$:
    \begin{lstlisting}
value = [1. 0.; 0. 1.]
    \end{lstlisting}
  \item
    \mc{ExpressionFunction} in \mc{dune/xt/functions/expression.hh}
models continuous functions, given an expression and order at runtime
(expressions for gradients can be optionally provided).
    Sample configuration for $f: \R^2 \to \R^2$ given by $(x, y) \mapsto (x,
\sin(y))$:
    \begin{lstlisting}
variable   = x
order      = 3
expression = [x[0] sin(x[1])]
gradient.0 = [1 0]
gradient.1 = [0 cos(x[1])]
    \end{lstlisting}
     Note that the user has to provide the approximation order, resulting in $f$
being locally approximated as a third order polynomial on each grid element.
  \item
    \mc{GlobalLambdaFunction} in \mc{dune/xt/functions/global.hh} models
continuous functions by evaluating a \Cpp{} lambda expression.
    Sample usage for $f: \R^2 \to \R$ given by $(x, y) \mapsto x$:
    \begin{c++}
GlobalLambdaFunction< E, D, 2, R, 1 > f([](DomainType x){ return x[0]; },
                                        1);  // <- local polynomial order
\end{c++}
  \item
    \mc{Spe10::Model1Function} in \mc{dune/xt/functions/spe10.hh} models
the permeability field of the SPE10 model1 test case, given the appropriate data
file\footnote{Available at \mc{http://www.spe.org/web/csp/datasets/set01.htm}}.
\end{itemize}

\bibliography{bibliography}
\end{document}